\documentclass[osajnl2,showpacs,twocolumn,superscriptaddress]{revtex4}
\usepackage{graphicx}
\usepackage{amsmath}

\begin{document}
\title{Singularities in Speckled Speckle}

\author{Isaac Freund}
\affiliation{Department of Physics, Bar-Ilan University, Ramat-Gan IL52900, ISRAEL}
\author{David A. Kessler}
\affiliation{Department of Physics, Bar-Ilan University, Ramat-Gan IL52900, ISRAEL}
\begin{abstract}
Speckle patterns produced by random optical fields with two (or more)
widely different correlation lengths exhibit speckle spots that are
themselves highly speckled.  Using computer simulations and analytic
theory we present results for the point singularities of speckled
speckle fields: optical vortices in scalar (one polarization component)
fields; C points in vector (two polarization component) fields.  In
single correlation length fields both types of singularities tend to be
more{}-or{}-less uniformly distributed.  In contrast, the singularity
structure of speckled speckle is anomalous: for some sets of source
parameters vortices and C points tend to form widely separated giant
clusters, for other parameter sets these singularities tend to form
chains that surround large empty regions.  The critical point
statistics of speckled speckle is also anomalous.  In scalar (vector)
single correlation length fields phase (azimuthal) extrema are always
outnumbered by vortices (C points).  In contrast, in speckled speckle
fields, phase extrema can outnumber vortices, and azimuthal extrema can
outnumber C points, by factors that can easily exceed
$10^{4}$ for experimentally realistic source parameters.\\
\end{abstract}
\maketitle

Studies of the singularities of random optical fields (speckle patterns)
normally involve fields with a single correlation length
${\Lambda}$~\cite{Goo07,Ber78,Hal81,Bar81,Liu92,Fre95,Rob96,Fre96,Fre98,Fre98a,Ber00,Ber01,Den02,Den03,Sos03,Fol03,Den03a,Wil04,Sos04,Ego05}. Here we study random fields with two
widely different ${\Lambda}$, finding anomalous spatial distributions
for the singularities, and anomalous statistics for them and their
associated critical points. These interesting anomalies can be
controlled by varying source parameters, and although studied here by
computer simulation and analytic theory, they should also be amenable
to experimental observation.

The presence of two distinct ${\Lambda}$ in a random optical field
is easily determined visually, because the major speckle
spots in the speckle pattern are themselves highly speckled, Figs. \ref{f1}
and \ref{f2}. We call such patterns ``speckled speckle'' ~\cite{F21}.

Speckled speckle, which arises from source distributions with widely
different length scales, is often the norm. It appears in an
uncontrolled form in many scattering experiments due to parasitic
scattering, but because the detector resolution is often insufficient
to resolve the small speckle spots, the presence of speckled speckle
may not be noticed. But it makes its presence felt in discrepancies
between experiment and theory, and is, for example, the most common
cause of the failure to obtain negative exponential statistics for the
intensity. Due to interference, parasitic fields with intensities
less than $10^{-8}$ of the main beam
can produce measurable effects.

Here we consider controlled situations in which the long correlation
length ${\Lambda}_a$ and the short correlation
length, ${\Lambda}_b$ and their associated
intensities $I_a$ and $I_b$, are chosen to have values that are
experimentally realistic.

Commonly used rotationally symmetric (isotropic) source distributions $S(r')$, and their associated autocorrelation functions $W(r)$,  for circular Gaussian fields ~\cite{Goo07} with a
single correlation length are: ({\em{i}}) a Gaussian,
${S}_{G}({r}') =
\exp[-{r}'^{2}/(2{p})^{2}]$,
${W}_{G}({r}) =
\exp(-{{\kappa}}^{2}{p}^{2}{r}^{2})$,
({\em{ii}}) a disk, ${S}_{D}({r}') =
{\Theta}({r}'-{p})$,
${W}_{D}({r}) =
2{J}_{1}({{\kappa}}{pr})/({{\kappa}}{pr})$,
and ({\em{iii}}) a thin ring, modeled as
${S}_{R}({r}') =
{\delta}({r}'{}-{p})$,
${W}_{R}({r}) =
{J}_{0}({{\kappa}}{pr})$. Here
${J}_{n}$ is a Bessel function of integer order
${n}$, and ${\kappa}=
2{\pi}/({\lambda}{Z})$, where ${\lambda}$ is the wavelength
of light, and ${Z}$ is the distance to the screen in the far field
on which the speckle pattern is measured~\cite{Goo07}.  In what follows we set
for convenience $\kappa= 1$. In each of the
above three cases ${\Lambda}\sim1/{p}$.

Controlled fields with two different correlation lengths and known
correlation functions are easily generated using linear combinations of
two of the above source functions with different values for the
parameter ${p}$. There are nine such combinations; here we
consider the three in which both source functions are the same {}-
two Gaussians, two disks, and two rings. Writing: ({\em{i}}) for
the Gaussians, ${S}_{a}({r}') =
\exp[-{r}'^{2}/(2{a})^{2}]$,
${W}_{a}({r}) =
\exp(-{a}^{2}{r}^{2})$,
${S}_{b}({r}') =
\exp[-{r}'^{2}/(2{b})^{2}]$,
${W}_{b}({r}) =
\exp(-{b}^{2}{r}^{2})$;
({\em{ii}}) for the disks, ${S}_{a}({r}') =
{\Theta}({r}'-{a})$,
${W}_{a}({r}) =
2{J}_{1}({ar})/({ar})$,
${S}_{b}({r}') =
{\Theta}({r}'-{b})$,
${W}_{b}({r}) =
2{J}_{1}({br})/({br})$; and ({\em{iii}}) for the
rings, ${S}_{a}({r}') =
{\delta}({r}'-{a})$,
${W}_{a}({r}) =
{J}_{0}({ar})$,
${S}_{b}({r}') =
{\delta}({r}'-{b})$,
${W}_{b}({r}) =
{J}_{0}({br})$; we have for our compound
source function ${S}_{ab}({r}')$ and its
scalar far{}-field autocorrelation function
${W}_{ab}({r})$,
\begin{subequations}
\begin{eqnarray}
{S}_{ab}({r}') &=&
{I}_{a}{S}_{a}({r}') +
{I}_{b}{S}_{b}({r}')    \\
{W}_{ab}({r}) &=&
[{W}_{a}({r}) +
{KW}_{b}({r})]/(1+{K}) \ ,
\end{eqnarray}
\end{subequations}
where for the Gaussians and the disks ${K} =
{b}^{2}{I}_{b}/({a}^{2}{I}_{a})$,
and for rings ${K}
={bI}_{b}/({aI}_{a})$. Here
${\Lambda}_{a} \sim 1/{a}$,
${\Lambda}_{b} \sim 1/{b}$.

 The point singularities of generic scalar (one polarization component)
fields are optical vortices~\cite{Ber78,Nye99}, whereas for generic vector (two
polarization component) fields the point singularities are C
points \textendash\  isolated points of circular polarization embedded in a field of
elliptical polarization~\cite{Ber01,Den02,Nye99}. It is convenient to treat C
points as vortices of the complex Stokes field
${S}_{12} = {S}_{1} +
i{S}_{2}$~\cite{Kon01,Fre01}, where in terms of Cartesian field
components ${E}_{x}$ and ${E}_{y}$,
and right (R) and left (L) circularly polarized components,
${E}_{R}$ and ${E}_{L}$, 
\begin{subequations}
\begin{eqnarray}
{S}_{1} &=&
|{E}_{x}{|}^{2}
-{|}{E}_{y}{|}^{2}
=
2Re({E}_{R}^{*}{E}_{L}), \\
{S}_{2} &=&
2Re({E}_{x}^{*}{E}_{y})
=
2Im({E}_{R}^{*}{E}_{L}).
\end{eqnarray}
\end{subequations}
Here Cartesian and circular field components are related by
\begin{equation}
{E}_{R} = ({E}_{x} -
i{E}_{y})/2^{1/2},\quad
{E}_{L} = ({E}_{x} +
i{E}_{y})/2^{1/2}
\end{equation}
C points are also optical vortices of ${E}_{R}$ and of
${E}_{L}$, because at a point where one of these
fields vanishes, as happens at an optical vortex, the optical
polarization is circular, whereas elsewhere in the field when both
components are nonzero the polarization is in general elliptical.
Thus the C point density is twice the vortex density.  Finally, for vortices, scalar fields ${a}$ and ${b}$ have the
same polarization, whereas for C points in vector fields ${a}$ and
${b}$ are orthogonally polarized.

These essential preliminaries completed, we turn now to our results for
speckled speckle in what may be called the perturbation regime, i.e.
the regime in which
${I}_{b}/{I}_{a}
\ll 1$. Results for other regimes will be reported
on separately.

In Fig. 1, which shows scalar fields, we compare normal speckle
(${I}_{b} = 0$) with speckled speckle
(${I}_{b} > 0$). As can be seen,
the vortices of speckled speckle are found only in regions surrounding
the ${a}$ field vortices shown in Fig. {\ref{f1}(b). The reason for this
is that at a vortex the total, i.e. composite, field amplitude must
vanish. But because ${I}_{b} \ll {I}_{a}$, the destructive
interference between the ${a}$ and ${b}$ fields required for
vortex generation can occur only in regions where
${I}_{a} = {I}_{b}$, i.e. where
${I}_{a}$ is very small. This leads to isolated
giant clusters of speckled speckle vortices near the positions of the
${a}$ field vortices. Such a cluster, containing 29 positive and
28 negative vortices is shown in Fig. \ref{f1}(e). As shown in Figs. \ref{f1}(f)
and \ref{f1}(g), vortices are absent in other regions.

An anomalous feature of speckled speckle phase maps is a granularity not
seen in normal speckle phase maps; not only the intensity, but also the
phase of speckled speckle is speckled. This granularity, which is
present throughout the phase map, and which can be seen most
easily in Fig. \ref{f1}(g), corresponds to large numbers of phase maxima and
minima, i.e. phase extrema.

\begin{figure}
\center{\includegraphics[width=0.45\textwidth]{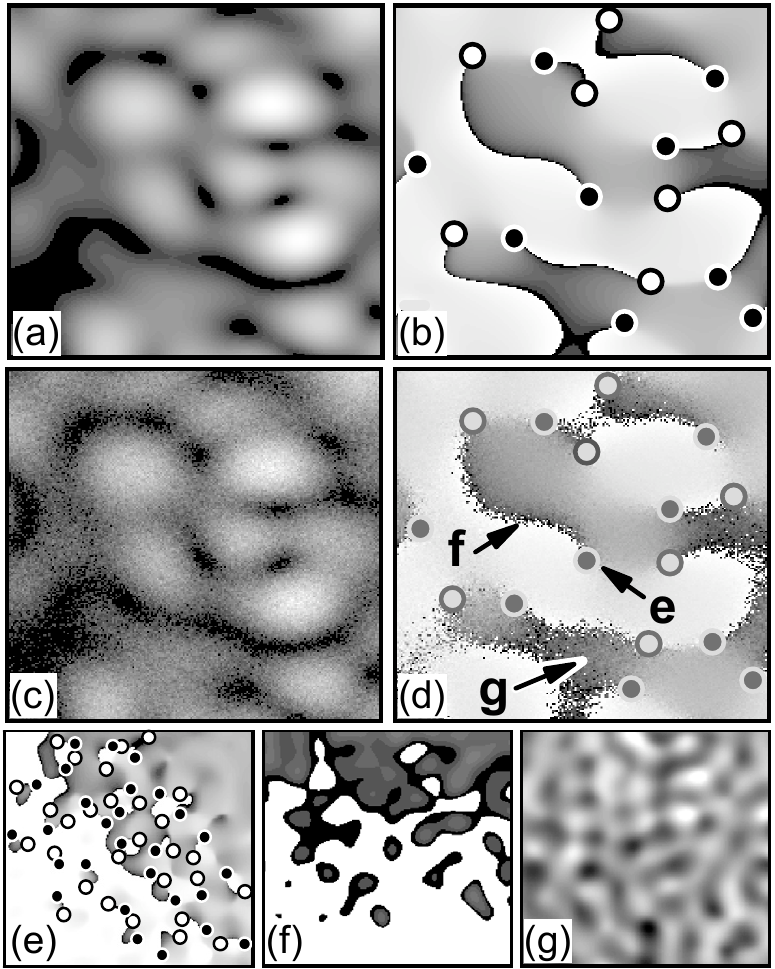}}
\caption{Scalar speckle. Shown are simulations for disks with
${a} = 1$, ${b }= 100$. In (a) and (b)
${I}_{b} = 0$, which yields familiar single
${\Lambda}$ speckle with ${\Lambda}\sim 1$. In (c)
{}- $(g) {I}_{b}/{I}_{a }=
2\cdot 10^{-6}$ (${K} = 0.02$),
which yields speckled speckle with
${\Lambda}_{a}\sim 1$,
${\Lambda}_{b}\sim 0.01$. (a) Intensity of
single ${\Lambda}$ speckle. (b) Phase map corresponding to (a).
(c) Intensity of speckled speckle. (d) Phase map corresponding to
(c). (e) {}- (f) Forty{}-fold enlargements of the regions in (d)
pointed to by the arrows labeled e, f, and g. Here and throughout
intensity (phase) is coded increasing (0 to 2${\pi}$) black to
white, and positive (negative) vortices are shown by filled white
(black) circles with black (white) rims.}
\label{f1}
\end{figure}

In normal speckle phase maps vortices $V$ always outnumber extrema $E$~\cite{Fre95,Den03}.  For a single Gaussian $E/V = 0.414$, for a single disk $E/V =
0.0887$, and for a single ring $E/V= 0$~\cite{Den03}. $E/V$ can be expressed in
terms of the moments $M_n$, $n = 2,4$, of the
source function $S$~\cite{Den03}.  For
speckled speckle we obtain for these moments ${M}_{n}
=
{C}_{n}({a}^{n}+{Kb}^{n})/(1+{K})$,
where for two Gaussians (2G), ${C}_{n }=
2^{n}(n/2)!$, for two disks (2D),
${C}_{n }= 2/(n+2)$, and for two rings (2R),
${C}_{n }= 1$. Writing ${\rho} =
{a}/{b}$, when ${K} =
{\rho}^{2}$,  $E/V$ attains a maximum value that for
small ${\rho}$ is given by
\begin{equation}
(E/V)_{max} =
{g}[(3)^{1/2}{\rho}^{2}]^{-1},
\end{equation}
where for 2G, ${g} = 1/2$, for 2D, ${g} = 1/3$, and for 2R,
${g} = 1/4$. In principle $(E/V)_{\textit{max}}$ can be made
arbitrarily large; for the parameters ${a}$ and ${b}$ in Fig.
1 we have for 2G, 2D, and 2R, respectively, $(E/V)_{\textit{max}} =
2887$, 1924, and 1443, corresponding to enhancements over normal speckle
by factors of $\sim 7,000$, $\sim 22,000$, and
${\infty}$.

\begin{figure}
\center{\includegraphics[width=0.45\textwidth]{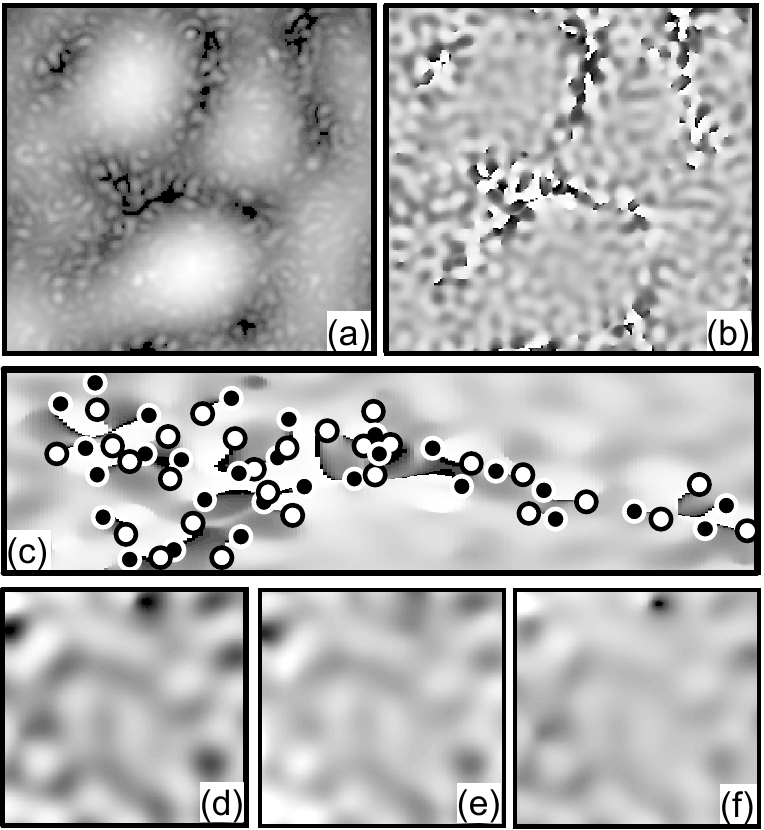}}
\caption{Vector field speckled speckle. Shown are simulations for disks
with {a} = 1, {b} = 10,
${I}_{b}/{I}_{a }=
3\cdot10^{-3}$ (${K} = 0.3$).
(a) Intensity. (b) Phase ${\Phi}_{12}$ of complex Stokes
field ${S}_{12}$. (c) Enlargement of the
black{}-white band on the horizontal centerline in (b). (d) \textendash\ (f)
Enlargement of the region beneath the black{}-white band in (b). (d)
Phase ${\Phi}_{12}$ of ${S}_{12}$. (e)
Phase $-{\Phi}_{R}$ of ${E}_{R}$.
(f) Phase ${\Phi}_{L }$ of ${E}_{L}$.}
\label{f2}
\end{figure}

 C points show similar anomalies. At a C point
${|}{E}_{x}{|} =
{|}{E}_{y}{|}$, so also speckled
speckle C points can be found only in dark regions of the {a}
field, here ${E}_{x}$. In Fig. 2 we show speckled
speckle C points for a value for
${I}_{b}/{I}_{a}$ that is much
larger than the corresponding value in Fig. \ref{f1}. Now C points are no
longer confined to the darkest regions near ${a}$ field vortices,
but can populate dark areas between
${a}$ field speckle spots, typically forming chains that tend to
surround these spots. An example of such a chain, which contains 28
positive and 29 negative C points, is shown in Fig. \ref{f2}(c).

 Also speckled speckle C point phase maps contain numerous extrema that
are absent in normal speckle C point maps: in obvious notation $E/C =
\frac12 E/V$. This result follows from the fact that, as shown
below, the density of phase extrema is essentially the same in both
scalar and vector speckled speckle, whereas the density of C points is
twice the vortex density~\cite{Ber01,Den02,Nye99}.

 From Eqs. (3)  we have ${{E}_R=
{E}_{x}(1-i{E}_{y}/{E}_{x})}$
  $\approx
{E}_{x}\exp(-i{E}_{y/}{E}_{x})$, 
$\quad{E}_{L }=
{{E}_{x}(1+i{E}_{y/}{E}_{x})}
\approx
{E}_{x}\exp(i{E}_{y}/{E}_{x})$,
where in the approximate equalities we have used the fact that
${E}_{x}$ is the strong ${a}$ field, and
${E}_{y}$ is the weak ${b}$ field, so that
${E}_{y/}{E}_{x}$ is small.
Writing ${S}_{12 }=
{A}_{12}\exp(i{\Phi}_{12})$,
${E}_{x}=
{A}_{x}\exp(i{\varphi}_{x})$,
${E}_{y}=
{A}_{y}\exp(i{\varphi}_{y})$,
${E}_{R }=
{A}_{R}\exp(i{\Phi}_{R})$, and
${E}_{L }=
{A}_{L}\exp(i{\Phi}_{L})$, and using
Eqs. (3), we have from Eqs. (2) ${\Phi}_{12 }=
{\Phi}_{L}-{\Phi}_{R} {\approx} 2({\Phi}_{L}-{\varphi}_{x})
{\approx}-2({\Phi}_{R}-{\varphi}_{x})$.
But because ${\Lambda}_{x} = {\Lambda}_{a
}\sim 1/{a}$ is large, except close to its sparse vortices
${\varphi}_{x}$ varies slowly compared to
${\Phi}_{12}$, ${\Phi}_{R}$, and
${\Phi}_{L}$, for which ${\Lambda}_{12}
{\approx}{\Lambda}_{R }{\approx}
{\Lambda}_{L }\sim 1/{b}$. Thus except close to
its sparse vortices ${\varphi}_{x}$ can be taken to be
nearly constant, and so the phase field ${\Phi}_{12}$ of
vector speckled speckle, and the phase fields ${\Phi}_{L}$
and $-{\Phi}_{R}$ of scalar speckled
speckle, have very nearly the same set of extrema.
This unusual property of speckled speckle is verified in Figs.
2(d){}-2(f).

In summary, we have shown that in the perturbation regime studied here
the singularities of scalar and vector speckled speckle, and the
critical points associated with these singularities, have highly
interesting, anomalous spatial arrangements and statistics. Singularity
screening, correlations, and other properties of these unusual fields
in different parameter regimes are also likely to exhibit unusual new
phenomena of special interest.

Email addresses: I. Freund, freund@mail.biu.ac.il; D. A. Kessler,
kessler@dave.ph.biu.ac.il.

\section*{Acknowledgements}
D. A. Kessler acknowledges the support of the Israel Science
Foundation.

\end{document}